\documentclass[conference, twocolumn]{IEEEtran}
\usepackage{cite}
\usepackage{amsmath,amssymb,amsfonts}
\usepackage{gensymb}
\usepackage{algorithmic}
\usepackage{graphicx}
\usepackage{textcomp}
\usepackage{xcolor}

\def\BibTeX{{\rm B\kern-.05em{\sc i\kern-.025em b}\kern-.08em
    T\kern-.1667em\lower.7ex\hbox{E}\kern-.125emX}}

\begin{document}

\title{Frequency Diverse Array OFDM Transmit System with Partial Overlap in Frequency}

\author{
    \IEEEauthorblockN{Marcin Wachowiak\IEEEauthorrefmark{1}\IEEEauthorrefmark{2}, 
    André Bourdoux\IEEEauthorrefmark{1}, 
    Sofie Pollin\IEEEauthorrefmark{2}\IEEEauthorrefmark{1}}
    \IEEEauthorblockN{
        \IEEEauthorrefmark{1} imec, Kapeldreef 75, 
        3001 Leuven, Belgium \\
        \IEEEauthorrefmark{2} Department of Electrical Engineering, 
        KU Leuven, Belgium \\
        Email: marcin.wachowiak@imec.be 
        }   
}

\maketitle

\begin{abstract}
A frequency-diverse array (FDA) is an alternative array architecture in which each antenna is preceded by a mixer instead of a phase shifter. The mixers introduce a frequency offset between signals transmitted by each antenna, resulting in a time-varying beam pattern. However, time-dependent beamforming is not desirable for communication or sensing. In this paper, the FDA is combined with orthogonal frequency-division multiplexing (OFDM) modulation. The proposed beamforming method partitions the OFDM symbol transmitted by all antennas into subcarrier blocks, which carry the same data but are precoded differently. The frequency offset between the antennas is equal to the subcarrier block width. Consequently, each antenna transmits a differently precoded subcarrier block at the center frequency, resulting in overlap and coherent summation of the blocks. Proposed architecture enables fully digital beamforming over a single block while requiring only a single digital-to-analog converter. The system's performance and tradeoffs are investigated in the context of joint communication and sensing.
\end{abstract}


\section{Introduction}

\subsection{Problem Statement}
Antenna arrays are envisaged to leverage the next generation of cellular networks - the 6G, by providing more degrees of freedom and beamforming gains to utilize the spectrum in the high-frequency bands \cite{6g_vision}. To meet the growing demands for flexible beamforming needed for spatial multiplexing, the antenna arrays are expected to scale up, resulting in a manifold increase in the number of radio front-end chips. To limit the cost and bring new capabilities, alternative antenna array architectures need to be considered \cite{alternative_array_arch}.
In the frequency-diverse array (FDA), each antenna is preceded by a mixer instead of a phase shifter. The mixer introduces a frequency offset across antennas that results in a time-varying phase difference. The time-dependent phase shift between antennas constitutes a time-varying beamforming vector. As a result, the FDA achieves the autoscanning property without the use of phase shifters. However, the FDAs have no dynamic control over their instantaneous beamforming direction, making them suitable for a limited range of applications.

\begin{figure}[t]
    \centering
    \includegraphics[width=\linewidth]{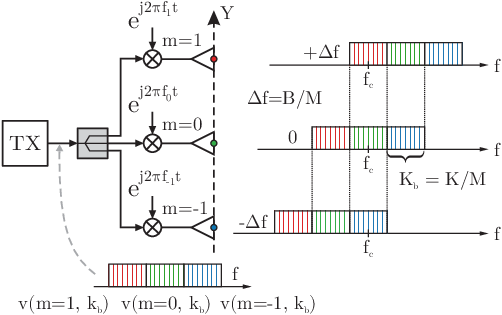}
    \caption{Proposed partially overlapped frequency diverse array and transmitted OFDM signals in the frequency domain. Illustrated for a case with M=3 antennas. All the antennas transmit the same baseband signal, but shifted in frequency depending on the antenna index $m$. The transmitted signal is partitioned into $M$ subcarrier blocks of size $K_{\mathrm{b}} = K/M$ carrying the same, repeated modulation symbols but precoded differently. The color denotes the precoder due to the spatial position of each antenna. Each of the subblocks is precoded for a different antenna position in such a way that the center block is digitally beamformed and experiences maximal array gain $M$.}
    \label{fig:fda_diagram}
\end{figure}

\subsection{Relevant works}
The concept of frequency diverse arrays was originally formulated for narrowband and continuous-wave radars \cite{antonik_fda_radar, tma_fda_dsp_overview, fda_techniques_for_radar}. The analyses did not consider wideband signals and investigated the beamforming only in the narrowband cases. As a result, the analyses neglected the overlap of the signals in the frequency domain and the resulting interference.
Possible ways of controlling the FDA beam pattern include random, non-uniform or time-modulated frequency offsets \cite{fda_beampattern_analysis}. These methods still produce a time-variant beam pattern with limited ways to control it.
The time-modulated FDAs \cite{time_modulated_fdas} can decouple the beam pattern in range and angle, resulting in pulsed phased array characteristics, which are not suitable for continuous wideband systems.
As the orthogonal division multiplexing (OFDM) modulation is considered for joint communication and sensing systems \cite{ofdm_jcs, mw_jcns_fda}, there are several works investigating the FDA OFDM systems.
In \cite{fda_ofdm_sc_per_antenna}, the OFDM FDA communications system is considered in the context of secure communications. The subcarriers of the OFDM symbol are partitioned between antennas and, as a result, the signal does not experience any array gain.
An integrated communications and sensing system using orthogonal chirps was presented in \cite{fda_ofdm_jcns_chirped}. The scheme also does not experience array gain as each chirp is transmitted by a separate antenna. This requires multiple RF front-ends and offers a low rate of communication. 
In \cite{fma_tx_jcns} the concept of a frequency diverse array was demonstrated in hardware. It investigated switched FDA and phased array operation. 
The narrowband FDA autoscanning mode was used only to detect the angle of the transmitting target.

\subsection{Contributions}

This work presents the OFDM FDA system with a novel method for achieving fully digital beamforming with a single digital-to-analog converter (DAC). The partition of the OFDM symbol into differently precoded blocks and adjusting the frequency offset enables fully digital beamforming over a single subcarrier block. The proposed partial overlap removes the undesirable autoscanning property at the expense of occupied bandwidth. The system's performance and trade-offs are assessed in the joint communication and sensing context. Moreover, methods of recycling the excessive bandwidth to improve the sensing and communication performance are proposed and evaluated.

\section{System model}

\subsection{Frequency diverse array}

Consider a uniform linear array (ULA) with $M$ elements. The indexing of the antennas is $m\in \{0, \pm1, \ldots, \pm(M-1)/2\}$ for odd $M$ and  $m\in \{\pm0.5, \pm1.5, \ldots, \pm(M-1)/2\}$ for even $M$. Each antenna transmits the same baseband signal at a distinct frequency given by 
\begin{equation}
    \label{eq:ant_freq_offset}
    f_m = f_c + m\Delta f,
\end{equation}
where $f_c$ is the center operating frequency and $\Delta f$ is the frequency offset between the antenna elements. 
The array is assumed to operate in the far field with the center of the array selected as the reference point. The path difference in the direction $\theta$ between the $m$th antenna and the reference is
\begin{equation}
    \label{eq:ula_path_diff}
    \Delta d_m = m d_{\lambda} \frac{c}{f_{\mathrm{c}}} \sin{(\theta)},
\end{equation}
where $d_{\lambda}$ is the spacing between antenna elements expressed in wavelengths
\subsection{Modulation}

The system utilizes OFDM modulation with $K$ subcarriers with symmetric indices $k \in \{-K/2, \ldots, K/2 - 1\}$. The subcarrier spacing is $\Delta f_{\mathrm{sc}}$ and the transmittter bandwidth is $B_{\mathrm{TX}} = K \Delta f_{\mathrm{sc}}$. The complex modulation symbols $s(k)$ are chosen from a QPSK constellation. 
The transmitted data symbol is partitioned into subcarrier blocks with repeated modulation symbols. The data symbols are repeated with periodicity $K/M$.
\begin{equation}
    \label{eq:data_symbols_rep}
    s(k) = s \left( k + m\frac{K}{M} \right).
\end{equation}
As can be seen, there are only $K_{\mathrm{b}}=K/M$ different modulation symbols forming $M$ subcarrier blocks with indices $b = m$. The subcarriers within a block have indices $k_{\mathrm{b}} \in \{-K_{\mathrm{b}}/2, \ldots, K_{\mathrm{b}}/2 - 1\}$.

\subsection{Precoding}

Next, each subcarrier block is precoded with a complex vector $v(b=m,k_{\mathrm{b}})$ that depends on block index $b$ and subcarrier index within the block $k_{\mathrm{b}}$. The precoding vector over all subcarriers of the OFDM symbol can be written as
\begin{equation}
    \label{eq:precoder_blocks}
    v(k) = v(b K_{\mathrm{b}} + k_{\mathrm{b}}) = v(b, k_{\mathrm{b}}).
\end{equation}
The precoded signal is then
\begin{equation}
    \label{eq:prec}
    x(k) = s(k) v(k).
\end{equation}
where, because of the periodicity of the modulation symbols $s(k) = s \left( k_{\mathrm{b}} + \lfloor \frac{k}{K_{\mathrm{b}}}\rfloor k_{\mathrm{b}} \right) \ \text{for} \ k>K_{\mathrm{b}}$.
Given the repeated symbol blocks and block precoding, the transmitted signal can be expressed in terms of subcarrier blocks
\begin{equation}
    \label{eq:prec_expanded}
    x(b, k_{\mathrm{b}}) = x(b K_{\mathrm{b}} + k_{\mathrm{b}}) = s(k_{\mathrm{b}}) v(b, k_{\mathrm{b}})
\end{equation}
The transmitted signal consists of subcarrier blocks, each of which is precoded for a different spatial position of the antenna. The $M$ subcarrier blocks are transmitted over $2M-1$ frequencies and with M antennas, in such a way that each antenna radiates the block with a precoder designed for its spatial position at the center frequency of the system.
Exactly one subcarrier block at the center frequency is digitally beamformed by all antennas (Fig. \ref{fig:fda_diagram}) in the direction of interest.

Moreover, the precoder, within the subcarrier block $K_{\mathrm{b}}$ can be partitioned up to create multiple simultaneous beams pointing in different directions. The multi-beam operation is achieved over different frequency resources - subcarriers within a block. The maximum number of simultaneous beams is equal to $K_{\mathrm{b}}$. The multi-beam operation is achieved at the cost of reduced communication and sensing performance due to reduced frequency resources.

\subsection{Frequency shift}

The time domain (TD) signal is obtained by performing an inverse discrete Fourier transform (IDFT) of the precoded symbols
\begin{equation}
    \label{eq:ofdm_mod}
    y(n) = \frac{1}{N} \sum_{k=0}^{N-1} x(k) e^{j2\pi\frac{n}{N}k}, 
\end{equation}
Each antenna transmits the same baseband signal but scaled and shifted in frequency, with the frequency offset being a function of the antenna index. The signal transmitted by $m$th antenna is
\begin{equation}
    \label{eq:freq_shift_ofdm_td}
    y_m(n) = \frac{1}{\sqrt{M}} y(n) e^{j2\pi m \frac{\Delta f}{N \Delta f_{\mathrm{sc}}} n } = \frac{1}{\sqrt{M}}y(n) e^{j2\pi m \frac{\Delta f}{B_{\mathrm{TX}}} n },
\end{equation}
where $1/\sqrt{M}$ normalizes the total transmit power to unity.
To facilitate the coherent summation of the subcarrier blocks and avoid intercarrier interference (ICI), the frequency offset is set to 
\begin{equation}
    \label{eq:freq_offset_partial_overlap}
    \Delta f = \frac{K}{M} \Delta f_{\mathrm{sc}} = \frac{B_{\mathrm{TX}}}{M}.
\end{equation}
The selected frequency offset introduces a partial overlap in between the signals transmitted by each antenna in the frequency domain. The coherent combination of differently precoded blocks across antennas facilitates digital precoding in the overlap region of $K_{\mathrm{b}}$ subcarriers. The selected frequency offset guarantees that all antennas share at least one subcarrier block,  which experiences full array gain. The combining of signals in the frequency domain is visualized in Fig. \ref{fig:fda_diagram}.
The total signal transmitted by the FDA consists of $2M-1$ subcarrier blocks, of which $M-1$ are shifted below and $M-1$ are shifted above the center frequency. The total bandwidth occupied by the partially overlapping FDA is $B = 2B_{\mathrm{TX}} - B_{\mathrm{TX}}/M$.

The frequency-domain representation of the signal transmitted by $m$th antenna is
\begin{equation}
    y_{m}(k_{\mathrm{f}}) = \frac{1}{\sqrt{M}} x\left(k - m K_{\mathrm{b}} \right),
\end{equation}
where $k_{\mathrm{f}} \in\left\{ -\frac{(M-1)}{2}K_{\mathrm{b}},\ ...,\ \frac{(M-1)}{2}K_{\mathrm{b}} \right\}$ is the subcarrier index of the total signal transmitted by the FDA.
Given the subcarrier block signal structure introduced in \eqref{eq:prec_expanded} the signal transmitted by each antenna can be rewritten as
\begin{equation}
    \label{eq:fd_shifted_tx_sig_block_form}
    y_{m}(b_{\mathrm{f}}) = \frac{1}{\sqrt{M}}x(b_{\mathrm{f}}-m, k_{\mathrm{b}}) 
    = \frac{1}{\sqrt{M}} s(k_{\mathrm{b}})v(b_{\mathrm{f}}-m, k_{\mathrm{b}}),
\end{equation}
where $b_{\mathrm{f}}$ is the block index of the total signal transmitted by the FDA with $b_{\mathrm{f}} \in \{0, \pm 1, \pm 2, \ldots, \pm (M - 1)\}$. The $v(b_{\mathrm{f}} - m)$ and thus $y_m(b_{\mathrm{f}})$ is defined and nonzero for $b_{\mathrm{f}} - m \in b$.
Due to frequency shift the precoder from $m$th antenna is shifted by $-m$ \eqref{eq:fd_shifted_tx_sig_block_form}. To match the precoder (index) to the antenna position the precoder of block $b$ is assigned to antenna $-m$ resulting in 
\begin{equation}
    \label{eq:prec_block_per_ant}
    v(b) = v(-m).
\end{equation}


\section{Signal model}
\label{sec:rx_sig_proc_model}

The communication and sensing channels share the same formulation. For a receiver/target (RT) located at angle $\theta_{\mathrm{RT}}$ and distance $d_{\chi}$ the channel is
\begin{equation}
    \label{eq:comm_los_channel_fda}
    h(m, b_{\mathrm{f}}, k_{\mathrm{b}}) = \chi e^{-j2 \pi \left( \frac{f_c + b_{\mathrm{f}} \Delta f + k_{\mathrm{b}}\Delta f_{\mathrm{sc}}}{c} \right) \left( d_{\chi} - m d_{\lambda} \frac{c}{f_{\mathrm{c}}} \sin{(\theta_{\mathrm{RT}})} \right) }.
\end{equation}
For communication, the distance $d_{\chi}$ is equal to $d$ and $\chi$ represents $\alpha$, the attenuation coefficient. For sensing $d_{\chi}$ is the bistatic distance $2d$ and $\chi$ is $\gamma$, the complex scattering coefficient. The sensing receiver is located in the center of the TX array.
The received signal over a single selected subcarrier block $b_{\mathrm{f}}$ is
\begin{align}
    \label{eq:rx_sig_per_block}
    r(b_{\mathrm{f}}, k_{\mathrm{b}}) &= s(k_{\mathrm{b}}) 
    \frac{1}{\sqrt{M}} \sum_{b_{\mathrm{f}} = b_{\mathrm{f, min}}}^{b_{\mathrm{f, max}}} 
    v(m - b_{\mathrm{f}}, k_{\mathrm{b}}) h(m, b_{\mathrm{f}}, k_{\mathrm{b}}).
\end{align}
where $b_{\mathrm{f, min}} = -(M-1)/2 + \max{(0, b_{\mathrm{f}})}$, and $b_{\mathrm{f, max}} = (M-1)/2 + \min{(0, b_{\mathrm{f}})}$.
The number of antennas (and blocks) overlapping in the frequency changes as $M - |b_{\mathrm{f}}|$. 
The gain experienced by each block of the total FDA signal is proportional to the number of overlapping blocks.
Depending on the block index $b_{\mathrm{f}}$, a different subset of antennas participates in the transmission of the block at a specific frequency. This creates an antenna subarray, which determines the beam pattern and phase center of the radiated block.
The greatest number of blocks equal to $M$ combines at $b_{\mathrm{f}}=0$, which is centered at $f_{\mathrm{c}}$.

\begin{figure}[t]
    \centering
    \includegraphics[width=\columnwidth]{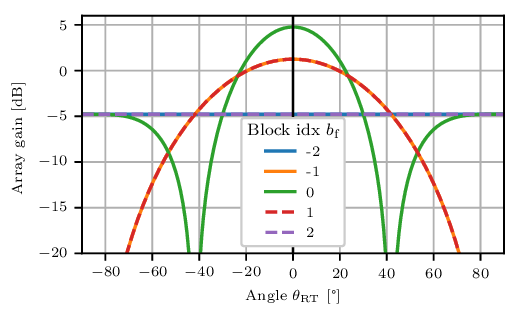}
    \caption{Array gain per subcarrier block of the total FDA signal for $M=3$ antennas and $\theta_{\mathrm{TX}} = 0\degree$.}
    \label{fig:beamforming_and_gain_per_block}
\end{figure}

Consider a constant precoder over the subcarrier block
\begin{equation}
    \label{eq:const_analog_precoder}
    v(b, k_{\mathrm{b}}) = e^{j2 \pi b (f_c + k_{\mathrm{b}} \Delta f_{\mathrm{sc}}) \frac{d_{\lambda}}{f_{\mathrm{c}}} \sin{(\theta_{\mathrm{TX}})}},
\end{equation}
where $b=-m$ can be used interchangeably and $\theta_{\mathrm{TX}}$ is the beamforming angle.
To simplify, the maximum frequency offset due to frequency shift and subcarrier spacing is considered negligible in regard to the center frequency $((2M-1)/2 )\Delta f + (K_{\mathrm{b}}/2) \Delta f_{\mathrm{sc}} \ll f_c$. Expanding \eqref{eq:rx_sig_per_block} with \eqref{eq:const_analog_precoder} and \eqref{eq:comm_los_channel_fda} gives
\begin{align}
    \label{eq:rx_sig_pb_full_form}
    r(b_{\mathrm{f}}, k_{\mathrm{b}}) = & s(k_{\mathrm{b}}) \beta(b_{\mathrm{f}}, k_{\mathrm{b}})
    e^{j2 \pi b_{\mathrm{f}} d_{\lambda} \sin{(\theta_{\mathrm{TX}})}} \nonumber \\ 
    & \times \frac{1}{\sqrt{M}} \sum_{b_{\mathrm{f}} = b_{\mathrm{f, min}}}^{b_{\mathrm{f, max}}}
    e^{j2 \pi m d_{\lambda} \left( \sin{(\theta_{\mathrm{RT}})} - \sin{(\theta_{\mathrm{TX}})} \right)},
\end{align}
where $\beta(b_{\mathrm{f}}, k_{\mathrm{b}}) = \chi e^{-j2 \pi \left( f_c + b_{\mathrm{f}} \Delta f + k_{\mathrm{b}}\Delta f_{\mathrm{sc}} \right)\frac{d_{\chi}}{c}}$ is a complex coefficient representing attenuation and phase rotation due to propagation. 
The sum in \eqref{eq:rx_sig_pb_full_form} can be interpreted as an array factor (AF) 
\begin{align}
    \label{eq:af_full_form}
    \mathrm{AF}(b_{\mathrm{f}},\theta_{\mathrm{RT}}, \theta_{\mathrm{TX}}) &= \frac{1}{\sqrt{M}} \sum_{b_{\mathrm{f}} = b_{\mathrm{f, min}}}^{b_{\mathrm{f, max}}}
    e^{j2 \pi m d_{\lambda} \left( \sin{(\theta_{\mathrm{RT}})} - \sin{(\theta_{\mathrm{TX}})} \right)}.
\end{align}
When the summation bounds are not symmetrical the array factor contains a phase shift component, which is due to the different positions of the phase center of the subarray and the array.
The AF can be separated into phase and amplitude components as follows
\begin{align}
    \label{eq:array_factor_subarray}
     \mathrm{AF}(b_{\mathrm{f}},\theta_{\mathrm{RT}}, \theta_{\mathrm{TX}}) =& 
     e^{j 2 \pi \frac{b_{\mathrm{f}}}{2} d_{\lambda} \left( \sin{(\theta_{\mathrm{RT}})} - \sin{(\theta_{\mathrm{TX}})} \right)} \nonumber \\  
     & \times \left| \mathrm{AF}( b_{\mathrm{f}},\theta_{\mathrm{RT}}, \theta_{\mathrm{TX}} ) \right|,
\end{align}
where $b_{\mathrm{f}} / 2$ can be interpreted as the subarray phase center index. 
By substituting \eqref{eq:array_factor_subarray} in \eqref{eq:rx_sig_pb_full_form} the received signal simplifies to
\begin{align}
    \label{eq:rx_sig_pb_simplified}
    r(b_{\mathrm{f}}, k_{\mathrm{b}}) = & s(k_{\mathrm{b}}) \beta(b_{\mathrm{f}}, k_{\mathrm{b}})
    e^{j \pi b_{\mathrm{f}} d_{\lambda}  \left( \sin{(\theta_{\mathrm{RT}})} + \sin{(\theta_{\mathrm{TX}})} \right)} \nonumber \\
    & \times \left| \mathrm{AF}( b_{\mathrm{f}},\theta_{\mathrm{RT}}, \theta_{\mathrm{TX}} ) \right|.
\end{align}
Due to the frequency shift, the precoder block indices are mismatched to the antenna indices (and the channel) for block indices other than $b_{\mathrm{f}} \neq 0$. The mismatch introduces an additional phase rotation, which is dependent on block index $b_{\mathrm{f}}$. The uniform antenna spacing guarantees linear phase progression across antennas and the precoder. The shift of the precoder indices in regard to antenna indices results in the same beamforming direction with additional phase rotation. As a result, all subcarrier blocks are beamformed in the same direction with the gain proportional to the number of overlapping blocks $M - |b_{\mathrm{f}}|$. 
Fig. \ref{fig:beamforming_and_gain_per_block} shows the AF per subcarrier block of the total FDA signal for $M=3$ antennas. The greater number of overlapping blocks improves the gain but reduces the half-power beamwidth.

In the previous derivations, the noise was omitted for simplicity. To account for it $n(b_{\mathrm{f}}, k_{\mathrm{b}})$ is added to \eqref{eq:rx_sig_pb_simplified}, which is the complex Gaussian noise per subcarrier within the block.

\subsection{Communication}

\subsubsection{Single block receiver}
Consider a narrowband receiver of bandwidth $B_{\mathrm{TX}}/M$ that observes a single, central subcarrier block $b_{\mathrm{f}} = 0$. 
In such a scenario, the received signal is
\begin{align}
    r(0, k_{\mathrm{b}}) = & s(k_{\mathrm{b}}) \beta(0, k_{\mathrm{b}})
    \left| \mathrm{AF}(0,\theta_{\mathrm{RT}}, \theta_{\mathrm{TX}} ) \right| + n(0, k_{\mathrm{b}}).
\end{align}
When the precoding and receiver/target angles match, the received signal simplifies to 
\begin{align}
    r(0, k_{\mathrm{b}}) = & s(k_{\mathrm{b}}) \sqrt{M} \beta(0, k_{\mathrm{b}}) + n(0, k_{\mathrm{b}}).
\end{align}
The gain of $M$ improves the signal-to-noise ratio (SNR). In the presence of beamforming angle mismatch  $\left( \sin{(\theta_{\mathrm{RT}}) -  \sin{(\theta_{\mathrm{TX}})}} \right)$, the gain is reduced accordingly. The attenuation and phase rotation $\beta(0, k_{\mathrm{b}})$ due to the propagation are compensated at the receiver.

\subsubsection{Full bandwidth receiver}

\begin{figure}[t]
    \centering
    \includegraphics[width=\columnwidth]{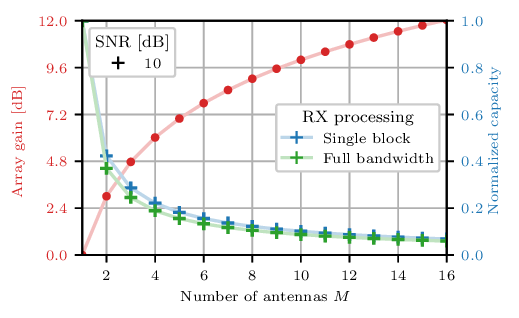}
    \caption{Array gain and normalized capacity of the system vs number of antennas for different receive processing and SNR.}
    \label{fig:fda_capacity_analysis}
\end{figure}

The total bandwidth occupied by the FDA ($B = 2B_{\mathrm{TX}} - B_{\mathrm{TX}}/M$) has limited use when considering communication performance. The full bandwidth is utilized by coherent summation of the $b_{\mathrm{f}}$ received subcarrier blocks. Given that the receiver has removed the effects of the channel by equalization the received communication signal $r_{\mathrm{c}}$ becomes
\begin{equation}
    r_{\mathrm{c}}(k_{\mathrm{b}}) = s(k_{\mathrm{b}})
    \sum_{b_{\mathrm{f}}=-(M-1)}^{M-1}  \left( \left| \mathrm{AF}( b_{\mathrm{f}},\theta_{\mathrm{RT}}, \theta_{\mathrm{TX}} ) \right| + n(b_{\mathrm{f}}, k_{\mathrm{b}}) \right).
\end{equation}
The power gain from the coherent summation of all the subcarrier blocks across all frequencies and antennas is equal to $M^2$.
However, due to wider bandwidth, the noise power also experiences gain equal to $2M - 1$.
As a result, the SNR gain from full FDA bandwidth processing is $G_{\mathrm{FB}} = M^2/(2M - 1) = M/(2 - 1/M)$, which is lower than for single-block processing $G_{\mathrm{SB}} = M$.
The lower SNR gain implies lower system capacity. However, the parallel channels can be exploited reduce the outage probability. Fig. \ref{fig:fda_capacity_analysis} presents the system normalized capacity as a function of the number of antennas for different receive processing and SNR = 10 dB.

\subsection{Sensing}

In the sensing application, the channel estimate $\hat{h}$ is obtained by zero-forcing as follows
\begin{equation}
    \label{eq:ch_est_general}
     \hat{h} (b_{\mathrm{f}}, k_{\mathrm{b}}) =  
     \frac{r(b_{\mathrm{f}}, k_{\mathrm{b}})}{s (k_{\mathrm{b}})} + \frac{n(b_{\mathrm{f}}, k_{\mathrm{b}})}{s (k_{\mathrm{b}}) }.
\end{equation}
The range profile is obtained by applying an IDFT over the channel estimate.

\subsubsection{Single block receiver}
\begin{figure}[t]
    \centering
    \includegraphics[width=\columnwidth]{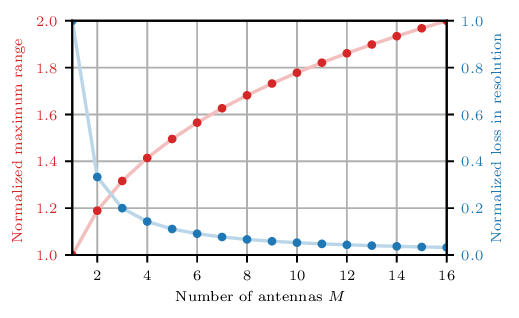}
    \caption{Maximum sensing range limited by path loss and range resolution normalized to a single antenna system as a function of the number of antennas.}
    \label{fig:fda_radar_analysis}
\end{figure}

When the beamforming and target angles match, the channel estimate of the $0$th block observed by the narrowband receiver becomes
\begin{equation}
    \label{eq:ch_est_center_sb}
     \hat{h} (0, k_{\mathrm{b}}) =  M \beta(0, k_{\mathrm{b}}) + \frac{n(0, k_{\mathrm{b}})}{s (k_{\mathrm{b}}) }.
\end{equation}
The maximum range of the considered system is limited by the length of the cyclic prefix (CP) due to communications functionality. The maximum unambiguous range and resolution for a single subcarrier block is
\begin{equation}
    \label{eq:max_range_res_single_block}
     r_{\mathrm{max}} = \frac{c N_{\mathrm{cp}} (2M - 1) }{2 B}, \quad r_{\mathrm{res}} = \frac{c (2M - 1)}{2B}.
\end{equation}
Given a fixed total FDA bandwidth $B$ the increase in the number of antennas reduces the range resolution obtained by a single block receiver. At the same time, the maximum achievable range at $b_{\mathrm{f}} = 0$ is improved due to beamforming gain. 
Fig. \ref{fig:fda_radar_analysis} illustrates the trade-off between the maximum range and sensing resolution as a function of the number of antennas. The results are normalized to a system with a single antenna.

\subsubsection{Full bandwidth receiver}
\label{sec:fb_sensing}

The use of other subcarrier blocks offers drastic improvements in the performance of the sensing system. The full bandwidth receiver can estimate the angle of the targets as each block is radiated by a different subarray. What is more, the obtained angular information can be used to compensate phase across different blocks to produce range profiles with refined resolution due to increased bandwidth.
When utilizing the full FDA-occupied bandwidth, the maximum range and resolution are identical to the standard OFDM system
\begin{equation}
    \label{eq:max_range_res_full_bw}
     r_{\mathrm{max}} = \frac{c N_{\mathrm{cp}}} {2 B}, \quad r_{\mathrm{res}} = \frac{c}{2 B}.
\end{equation}
Given the information about target angles, the $2M - 1$ times increase in range resolution and maximum unambiguous range is achieved at the cost of a $2M - 1$ increase in RX ADC bandwidth. The maximum unambiguous Doppler velocity and resolution remain the same as in a conventional OFDM system. 

To perform angle of arrival (AoA) estimation the received signal needs to be compensated to remove the effects of transmit precoding which are phase shift and per block power variations (windowing). The channel estimate used for the angle of arrival estimation is
\begin{align}
    \hat{h}_{\mathrm{AoA}}(b_{\mathrm{f}}, k_{\mathrm{b}}) =& 
    \frac{\hat{h} (b_{\mathrm{f}}, k_{\mathrm{b}})}{\left| \hat{h} (b_{\mathrm{f}}, k_{\mathrm{b}})\right|} 
    e^{-j \pi b_{\mathrm{f}} d_{\lambda} \sin{(\theta_{\mathrm{TX}})}} \\
    = & \arg{ \left( \beta(b_{\mathrm{f}}, k_{\mathrm{b}}) e^{j \pi b_{\mathrm{f}} d_{\lambda} \sin{(\theta_{\mathrm{RT}})}} + 
    \frac{n(b_{\mathrm{f}}, k_{\mathrm{b}})}{s(k_{\mathrm{b}})} \right) } \nonumber.
\end{align}
The AoA estimation can be performed by a matched filter or DFT across the range profiles obtained from each of the blocks. For DFT AoA the range profiles of each block need to be compensated for the range-angle coupling as their center frequency are different. The resolution of the angle estimation is identical to that of conventional methods with $M$ antennas. The range resolution of the obtained range-angle map is determined by the bandwidth of the single subcarrier block.

\begin{figure}[t]
    \centering
    \includegraphics[width=\columnwidth]{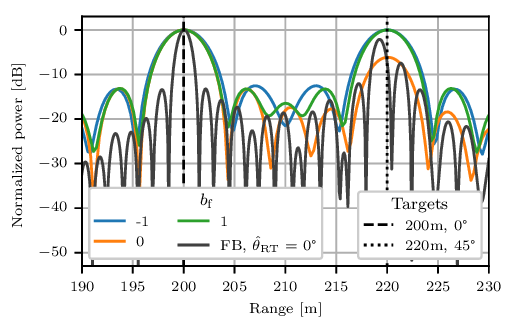}
    \caption{Range profile per subcarrier block and full bandwidth (FB) processing for a system with $M=2$ antennas and 2 targets. The array is beamforming at the boresight $\theta_{\mathrm{TX}} = 0\degree$. The AoA used in full bandwidth processing corresponds to the first target angle $\hat{\theta}_{\mathrm{RT}} = 0\degree$. The remaining parameters are provided in Sec. \ref{sec:sim_results}. }
    \label{fig:fda_rps}
\end{figure}

Once the estimate of target angle $\hat{\theta}_{\mathrm{RT}}$ is obtained, it can be used to obtain a refined resolution range profile. In order to coherently process all subcarrier blocks the received signal is divided by the estimated array factor to remove the undesired windowing and each block is phase-compensated. The equalized channel estimate is obtained as follows
\begin{align}
    \label{eq:ch_est_fb_eq}
    \hat{h}_{\mathrm{EQ}}(b_{\mathrm{f}}, k_{\mathrm{b}}) =& \frac{\hat{h}(b_{\mathrm{f}}, k_{\mathrm{b}})}{\left| \mathrm{AF}( b_{\mathrm{f}},\hat{\theta}_{\mathrm{RT}}, \theta_{\mathrm{TX}} ) \right|}
    e^{-j \pi b_{\mathrm{f}} d_{\lambda}  \left( \sin{(\hat{\theta}_{\mathrm{RT}})} + \sin{(\theta_{\mathrm{TX}})} \right)} \nonumber \\
    =& \beta(b_{\mathrm{f}}, k_{\mathrm{b}})
    \frac{\left| \mathrm{AF}( b_{\mathrm{f}},\theta_{\mathrm{RT}}, \theta_{\mathrm{TX}} ) \right|}{
    \left| \mathrm{AF}( b_{\mathrm{f}},\hat{\theta}_{\mathrm{RT}}, \theta_{\mathrm{TX}} ) \right|
    } \nonumber \\
    & \times e^{j \pi b_{\mathrm{f}} d_{\lambda}  \left( \sin{(\theta_{\mathrm{RT}})} - \sin{(\hat{\theta}_{\mathrm{RT}})} \right)}.
\end{align}
Finally, to obtain a full bandwidth range profile, IDFT is performed over all compensated subcarrier blocks. Fig. \ref{fig:fda_rps} shows the range profile per subcarrier block and full bandwidth processing for $M=2$ antennas and 2 targets. Considered full bandwidth processing \eqref{eq:ch_est_fb_eq} is based on a single target AoA. In the presence of multiple targets at other angles, the mismatched phase and amplitude equalization results in a biased range estimate and increased sidelobes. Therefore, the full bandwidth range profile has to be calculated for each estimated target angle. The presence of multiple targets does not affect the performance when a single block sensing is considered.
 
\section{Simulation Results}
\label{sec:sim_results}
To assess the system's performance a series of simulations was performed with the following parameters; center frequency $f_{\mathrm{c}} = 28$ GHz, total FDA bandwidth $B=100$ MHz, number of subcarriers $N_{\mathrm{sc}} = 1024$ and number of antennas $M= \{ 2,4,8,16 \}$. The total FDA bandwidth is constrained to $B$, therefore the sampling frequency of the DAC is dependent on the number of antennas $f_\mathrm{s} = B M / (2M-1)$.
The sensing performance under imperfections is assessed by the integrated sidelobe level (ISL)
\begin{equation}
    \label{eq:isl_definition}
    \mathrm{ISL} = \frac{{\sum_{k_{\mathrm{f}}=0, \  k_{\mathrm{f}} \neq k_{{\mathrm{f}}_{max}} }^{K-1} \lvert p(k_{\mathrm{f}}) \rvert }^{2} }{|p(k_{{\mathrm{f}}_{max}})|^2},
\end{equation}
where $p(k_{\mathrm{f}})$ is the range profile obtained by full bandwidth processing of \eqref{eq:ch_est_fb_eq} and $k_{{\mathrm{f}}_{max}}$ is the index of the maximum of the range profile.

\begin{figure}[t]
    \centering
    \includegraphics[width=\columnwidth]{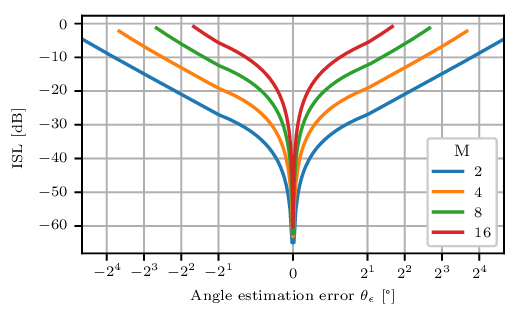}
    \caption{ISL of the range profile obtained from full bandwidth processing as a function of the angle estimation error for a single target and selected number of antennas $M$.}
    \label{fig:isl_evm_vs_aerr}
\end{figure}

\subsection{Angle estimation error sensitivity}
As introduced in Sec. \ref{sec:fb_sensing}, the angle estimation error introduces mismatched equalization in \eqref{eq:ch_est_fb_eq}, which gives rise to increased sidelobes and degraded ISL of the full bandwidth range profile. The angle estimation error is defined as $\theta_{\epsilon} = \hat{\theta}_{\mathrm{RT}} - \theta_{\mathrm{RT}}$. In the simulations, the target is located at boresight $\theta_{\mathrm{RT}} = 0\degree$ and the estimation error is kept within half the power beamwidth of the antenna array.
Fig. \ref{fig:isl_evm_vs_aerr} presents the effect of the angle estimation error on the ISL of the full bandwidth range profile. As can be observed, the impact of the error is more severe for a greater number of antennas due to the multiplicative nature of the error. The system is relatively sensitive to angular error and requires the error to be smaller than $2 \degree$ for most cases.

\begin{figure}[t]
    \centering
    \includegraphics[width=\columnwidth]{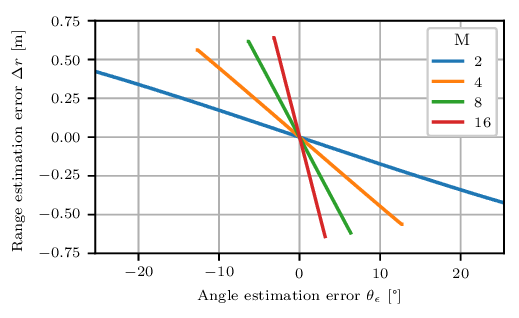}
    \caption{Range estimation error as a function of angle estimation error for a single target located at the boresight $\theta_{\mathrm{RT}} = 0\degree$ and selected number of antennas $M$.}
    \label{fig:rerr_vs_aerr}
\end{figure}

The phase of the full bandwidth channel estimate from \eqref{eq:ch_est_fb_eq} is a function of the estimated target angle $\hat{\theta}_{\mathrm{RT}}$. 
As a result, the range profile obtained from full bandwidth processing is range-angle coupled. Due to range-angle coupling, the angle estimation error introduces a range estimation error. 
Fig. \ref{fig:rerr_vs_aerr} presents the range estimation error as a function of the angle estimation error. For the target at boresight, the range error can be approximated as $\Delta r = -  \frac{1}{4} \frac{c}{\Delta f} d_\lambda \sin{ \left( \theta_{\epsilon} \right) }$. For a fixed total FDA bandwidth, the increase in the number of antennas reduces the frequency spacing as in \eqref{eq:freq_offset_partial_overlap}, resulting in increased range estimation error.

\section{Conclusion}
In this paper, a partially overlapped OFDM FDA system is proposed and analyzed. The system achieves digital beamforming over a subset of subcarriers by using a single DAC and $M$ RF mixers. As compared to conventional FDA in this system, the beam pattern is no longer time-varying due to the utilized repeated subcarrier block structure and precoding. The beamforming gain is achieved at the cost of approximately twice the bandwidth overhead. The full bandwidth of the considered system can be leveraged to estimate the target angles and refine the range resolution. The proposed hardware architecture offers a cost and energy-effective solution for digital beamforming.

\bibliographystyle{IEEEtran}
\bibliography{biblio.bib}

@ARTICLE{6g_vision,
  author={Tataria, Harsh and Shafi, Mansoor and Molisch, Andreas F. and Dohler, Mischa and Sjöland, Henrik and Tufvesson, Fredrik},
  journal={Proceedings of the IEEE}, 
  title={{6G} Wireless Systems: Vision, Requirements, Challenges, Insights, and Opportunities}, 
  year={2021},
  volume={109},
  number={7},
  pages={1166-1199},
  doi={10.1109/JPROC.2021.3061701}
}

@INPROCEEDINGS{antonik_fda_radar,
  author={Antonik, P. and Wicks, M.C. and Griffiths, H.D. and Baker, C.J.},
  booktitle={2006 IEEE Conference on Radar}, 
  title={Frequency diverse array radars}, 
  year={2006},
  volume={},
  number={},
  pages={3 pp.-},
  doi={10.1109/RADAR.2006.1631800}
}

@ARTICLE{tma_fda_dsp_overview,
  author={Wang, Wen-Qin and So, Hing Cheung and Farina, Alfonso},
  journal={IEEE Journal of Selected Topics in Signal Processing}, 
  title={An Overview on Time/Frequency Modulated Array Processing}, 
  year={2017},
  volume={11},
  number={2},
  pages={228-246},
  doi={10.1109/JSTSP.2016.2627182}
}

@ARTICLE{fda_techniques_for_radar,
  author={Sammartino, Pier Francesco and Baker, Christopher J. and Griffiths, Hugh D.},
  journal={IEEE Transactions on Aerospace and Electronic Systems}, 
  title={Frequency Diverse {MIMO} Techniques for Radar}, 
  year={2013},
  volume={49},
  number={1},
  pages={201-222},
  doi={10.1109/TAES.2013.6404099}
}

@article{fda_ofdm_sc_per_antenna,
author = {Ding, Y. and Zhang, J. and Fusco, V.},
title = {Frequency diverse array {OFDM} transmitter for secure wireless communication},
journal = {Electronics Letters},
volume = {51},
number = {17},
pages = {1374-1376},
doi = {https://doi.org/10.1049/el.2015.1491},
year = {2015}
}

@ARTICLE{fda_ofdm_jcns_chirped,
  author={Huang, He and Wang, Wen-Qin},
  journal={IEEE Aerospace and Electronic Systems Magazine}, 
  title={{FDA-OFDM} for integrated navigation, sensing, and communication systems}, 
  year={2018},
  volume={33},
  number={5-6},
  pages={34-42},
  doi={10.1109/MAES.2018.170109}}

@ARTICLE{alternative_array_arch,
  author={Rocca, Paolo and Oliveri, Giacomo and Mailloux, Robert J. and Massa, Andrea},
  journal={Proceedings of the IEEE}, 
  title={Unconventional Phased Array Architectures and Design Methodologies—A Review}, 
  year={2016},
  volume={104},
  number={3},
  pages={544-560},
  doi={10.1109/JPROC.2015.2512389}}

@ARTICLE{time_modulated_fdas,
  author={Chen, Kejin and Yang, Shiwen and Chen, Yikai and Qu, Shi-Wei},
  journal={IEEE Transactions on Antennas and Propagation}, 
  title={Accurate Models of Time-Invariant Beampatterns for Frequency Diverse Arrays}, 
  year={2019},
  volume={67},
  number={5},
  pages={3022-3029},
  doi={10.1109/TAP.2019.2896712}}

@ARTICLE{fda_beampattern_analysis,
    author={Ahmad, Zeeshan and Chen, Meng and Bao, Shu-Di},
    title={Beampattern analysis of frequency diverse array radar: a review},
    journal={EURASIP Journal on Wireless Communications and Networking},
    year={2021},
    month={Nov},
    day={16},
    volume={2021},
    number={1},
    pages={189},
    issn={1687-1499},
    doi={10.1186/s13638-021-02063-6},
}

@ARTICLE{fma_tx_jcns,
  author={Mannem, Naga Sasikanth and Erfani, Elham and Huang, Tzu-Yuan and Wang, Hua},
  journal={IEEE Journal of Solid-State Circuits}, 
  title={A mm-Wave Frequency Modulated Transmitter Array for Superior Resolution in Angular Localization Supporting Low-Latency Joint Communication and Sensing}, 
  year={2023},
  volume={58},
  number={6},
  pages={1572-1585},
  doi={10.1109/JSSC.2022.3212207}}

@ARTICLE{ofdm_jcs,
  author={Sturm, Christian and Wiesbeck, Werner},
  journal={Proceedings of the IEEE}, 
  title={Waveform Design and Signal Processing Aspects for Fusion of Wireless Communications and Radar Sensing}, 
  year={2011},
  volume={99},
  number={7},
  pages={1236-1259},
  doi={10.1109/JPROC.2011.2131110}}

@INPROCEEDINGS{mw_jcns_fda,
  author={Wachowiak, Marcin and Bourdoux, André and Pollin, Sofie},
  booktitle={2025 IEEE 5th International Symposium on Joint Communications \& Sensing (JC\&S)}, 
  title={Frequency Diverse Array OFDM System for Joint Communication and Sensing}, 
  year={2025},
  volume={},
  number={},
  pages={1-6},
  doi={10.1109/JCS64661.2025.10880642}}

\end{document}